\documentclass{webofc}

\usepackage[varg]{txfonts}   % Web of Conferences font
\usepackage{hyperref}

\def\ba{\begin{eqnarray}}
\def\ea{\end{eqnarray}}
\def\be{\begin{equation}}
\def\ee{\end{equation}}

\begin{document}

\title{Bottomonium suppression and flow in heavy-ion collisions}

\author{
Michael Strickland \inst{1}\fnsep\thanks{\email{mstrick6@kent.edu}}
}

\institute{Department of Physics, Kent State University, Kent, OH 44242}

\abstract{
The strong suppression of bottomonia production in ultra-relativistic heavy-ion collisions is a smoking gun for the creation of a deconfined quark-gluon plasma (QGP).  In this proceedings contribution, I review recent work that aims to provide a more comprehensive and systematic understanding of bottomonium dynamics in the QGP through the use of pNRQCD and an open quantum systems approach.  This approach allows one to evolve the heavy-quarkonium reduced density matrix, taking into account non-unitary effective Hamiltonian evolution of the wave-function and quantum jumps between different angular momentum and color states.  In the case of a strong coupled QGP in which $E_{\rm bind} \ll T,m_D \ll 1/a_0$, the corresponding evolution equation is Markovian and can therefore be mapped to a Lindblad evolution equation.  To solve the resulting Lindblad equation, we make use of a stochastic unraveling called the quantum trajectories algorithm and couple the non-abelian quantum evolution to a realistic 3+1D viscous hydrodynamical background.  Using a large number of Monte-Carlo sampled bottomonium trajectories, we make predictions for bottomonium $R_{AA}$ and elliptic flow as a function of centrality and transverse momentum and compare to data collected by the ALICE, ATLAS, and CMS collaborations.
}

\maketitle

\section{Introduction}
\label{intro}

Bottomonium states that propagate through a deconfined quark-gluon plasma (QGP) are suppressed due to both Debye-screening, which modifies the real part of the heavy-quark potential, and in-medium transitions/breakup, which are encoded in the imaginary part of the heavy-quark potential.  In recent years there have significant advances in our understanding of heavy-quark bound state dynamics in the QGP stemming from a combination of potential non-relativistic QCD (pNRQCD) and open quantum systems (OQS) \cite{Brambilla:2016wgg,Brambilla:2017zei,Brambilla:2020qwo,Brambilla:2021wkt}.  In this effective field theory treatment one relies on a separation scales between the inverse size of the bound states $1/\langle r \rangle$, the effective temperature of the system $T$, and the binding energy of the states $E$.  Through the inclusion of hard-thermal-loop effects one can also include the effect of the induced Debye mass, $m_D$, of the system on the heavy-quark potential.  In a strongly coupled QGP, due to the large coupling, there is not a strict ordering of the temperature and Debye mass scales.  Instead one typically has $\pi T \sim m_D$ and the relevant hierarchy of scales is $1/a_0 \gg \pi T, m_D \gg  E$, where $a_0$ is the Bohr radius.  In this case it can be shown that the quantum evolution is Markovian, with the local medium relaxation time being shorter than both the time scale associated with internal transitions and the probe (heavy quarkonium) relaxation time \cite{Akamatsu:2020ypb,Rothkopf:2020vfz,Yao:2021lus}.  As a consequence, the dynamical equation which governs the evolution of the heavy quarkonium reduced density matrix is of Lindblad form \cite{Brambilla:2016wgg,Brambilla:2017zei}.

The Lindblad equation for the reduced density is challenging to solve numerically.  This stems from the fact that one must decompose states in angular momentum and color quantum numbers followed by discretization of the underlying wave-functions on a lattice of size $N$, resulting in a reduced density matrix with the number of elements proportional to $(l_\text{max}+1)^2 N^2$, where $l_\text{max}$ is the largest angular momentum quantum number considered.  For reliable computation one must take both a large number of lattice points and a large angular momentum cutoff, which makes directly solving the Lindblad equation numerically prohibitive, with memory size scaling like $(l_\text{max}+1)^2 N^2$ and per step evolution times scaling like $(l_\text{max}+1)^4 N^4$ \cite{Omar:2021kra}.\footnote{For application to heavy-quarkonium dynamics in the QGP one needs $N \sim 4000$ and $l_\text{max} \sim 5$.}  When faced with a problem with such high dimensionality it is frequently beneficial to make use of Monte-Carlo methods.  A particularly well-suited algorithm is provided by the quantum trajectories algorithm \cite{Dalibard:1992zz,Molmer:93,Plenio:1997ep,carmichael1999statistical,weissbook,Daley:2014fha}.  In this approach, one maps the solution of the Lindblad equation to solution of an ensemble of a independent one-dimensional Schr\"odinger equation evolutions that are subject to stochastic color and angular momentum transitions (quantum jumps) \cite{Daley:2014fha}.  When averaged over the ensemble, one obtains a solution to the three-dimensional Lindblad equation.  Due to the fact that each quantum trajectory is independent, the quantum trajectories algorithm lends itself to massive parallel computation.  Additionally, since between quantum jumps the evolution proceeds with fixed color and angular momentum quantum numbers, one does not have to place a cutoff on the magnitude of the angular momentum.  Herein, we present results obtained using an open-source code called QTraj, which implements the quantum trajectories algorithm for bottomonium states \cite{Omar:2021kra}.

In this proceedings contribution, we briefly review the theoretical underpinnings of the method and present comparisons between the phenomenological results obtained using a realistic hydrodynamical background and experimental data for bottomonium observables collected by the ALICE, ATLAS, and CMS collaborations in 5 TeV Pb-Pb collisions.

\section{Methodology}
\label{sec:methods}

Using pNRQCD and OQS, in Refs. \cite{Brambilla:2016wgg,Brambilla:2017zei} the authors obtained a set of master equations for heavy quarkonium in a strongly coupled QGP.  When the temperature is much larger than the binding energy, $T \gg E$, one can make an expansion in $E/T$, finding at leading order an evolution equation of Lindblad form~\cite{Lindblad:1975ef,Gorini:1975nb}
\begin{equation}\label{eq:lindblad}
\frac{d \rho(t)}{dt} = -i[H, \rho(t)] + \sum_{n} \left( C_{n} \rho(t) C_{n}^{\dagger} - \frac{1}{2} \left\{ C_{n}^{\dagger}C_{n}, \rho(t) \right\} \right),
\end{equation}
where, for $N_c$ colors, the reduced density matrix and Hamiltonian are given by
\begin{align}
\rho(t) =& \begin{pmatrix} \rho_{s}(t) & 0 \\ 0 & \rho_{o}(t) \end{pmatrix},\\
H =& \begin{pmatrix} h_{s} & 0 \\ 0 & h_{o} \end{pmatrix} + \frac{r^{2}}{2} \gamma \begin{pmatrix} 1 & 0 \\ 0 & \frac{N_{c}^{2}-2}{2(N_{c}^{2}-1)} \end{pmatrix} ,
\end{align}
with $\rho_{s}(t)$ and $\rho_{o}(t)$ being the singlet and octet reduced density matrices, respectively, and $h_{s,o}={\mathbf{p}^{2}}/{M} + V_{s,o}$ being the singlet or octet Hamiltonians.
The jump (collapse) operators $C$ appearing in Eq.~\eqref{eq:lindblad} are 
\be
C_{i}^{0} = \sqrt{\frac{\kappa}{N_{c}^{2}-1}} r^{i} \begin{pmatrix} 0 & 1 \\ \sqrt{N_{c}^{2}-1} & 0 \end{pmatrix},  \hspace{5mm}
C_{i}^{1} = \sqrt{\frac{(N_{c}^{2}-4)\kappa}{2(N_{c}^{2}-1)}} r^{i} \begin{pmatrix} 0 & 0 \\ 0 & 1 \end{pmatrix},
\ee
with the transport coefficients $\kappa$ and $\gamma$ given by the real and imaginary parts of the chromoelectric correlator
\be
\kappa = \frac{g^{2}}{18} \int_{0}^{\infty} dt \left\langle \left\{ \tilde{E}^{a,i}(t,\mathbf{0}), \tilde{E}^{a,i}(0,\mathbf{0}) \right\} \right\rangle, \hspace{5mm}
\gamma = -i \frac{g^{2}}{18} \int_{0}^{\infty} dt \left\langle \left[ \tilde{E}^{a,i}(t,\mathbf{0}), \tilde{E}^{a,i}(0,\mathbf{0}) \right] \right\rangle  .
\ee

\section{Results}
\label{sec:results}

We solve \eqref{eq:lindblad} using the quantum trajectories algorithm.  For the transport coefficients $\kappa$ and $\gamma$, we use lattice quantum chromodynamics measurements to constrain them.  For the results presented herein we will vary $\kappa$ over three temperature-dependent parameterizations $\hat\kappa(T) = \kappa(T)/T^3 \in \{ \hat\kappa_L(T), \hat\kappa_C(T),\hat\kappa_U(T)\}$ with the lower, central, and upper limits determined using the results of Ref.~\cite{Brambilla:2020siz}.  For $\gamma$, less is known.  To allow for all current indirect lattice extractions, we take $\hat{\gamma} = \gamma / T^{3} = \{-3.5,\, -1.75, \, 0 \}$ \cite{Brambilla:2019tpt,Kim:2018yhk,Aarts:2011sm,Larsen:2019bwy,Shi:2021qri}.  We note that the coefficient $\kappa$ sets the magnitude of the imaginary part of the potential, while $\gamma$ sets the magnitude of the modification of the real part of the potential.

%%%%%%%%%%%%%%%%%%%%%%%%%%%%%%%%%%%%%%%%%%%%%%%%%%%%%%%%%%%
\begin{figure}[t!]
\centerline{
\includegraphics[width=0.465\linewidth]{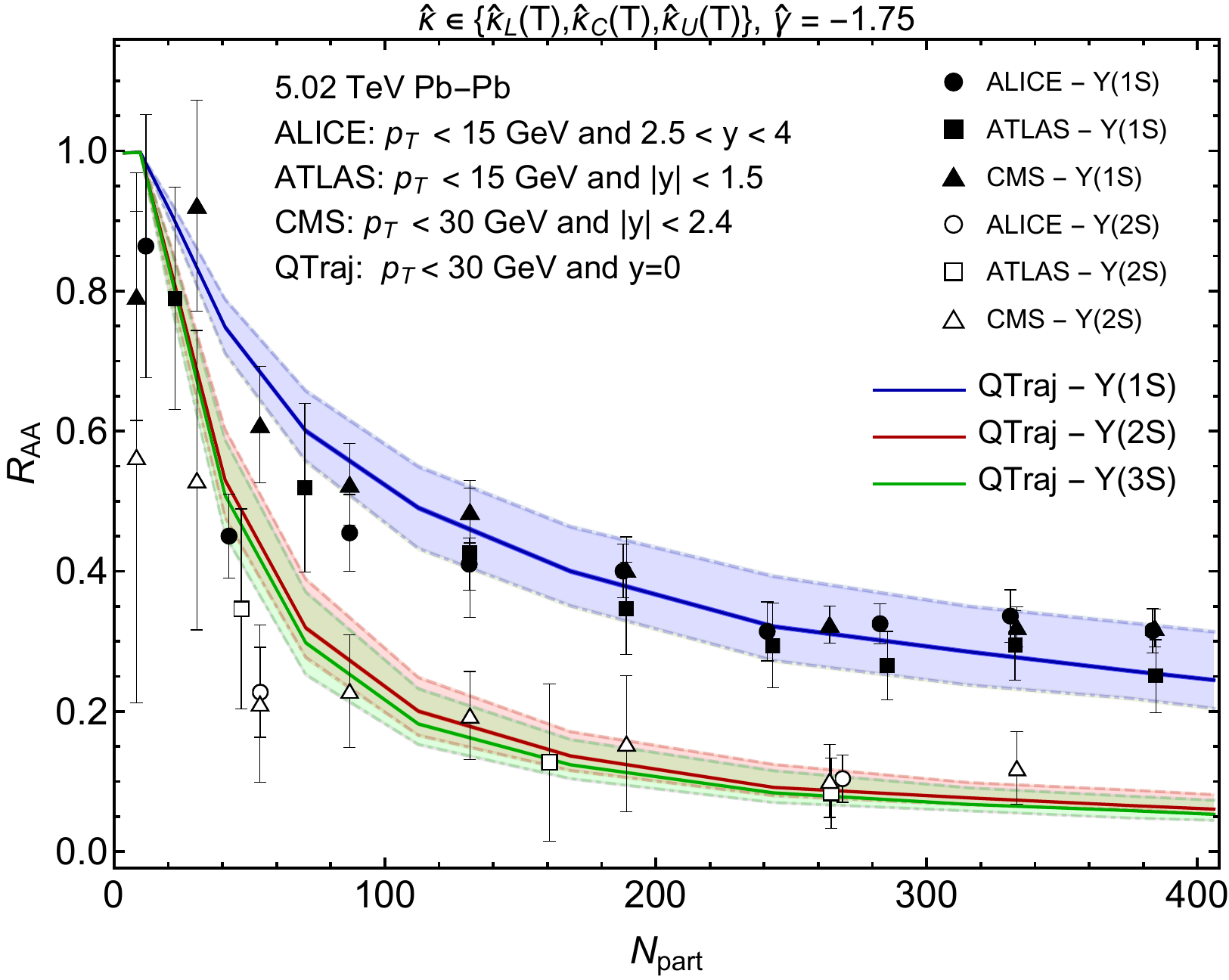}\hspace{2mm}
\includegraphics[width=0.465\linewidth]{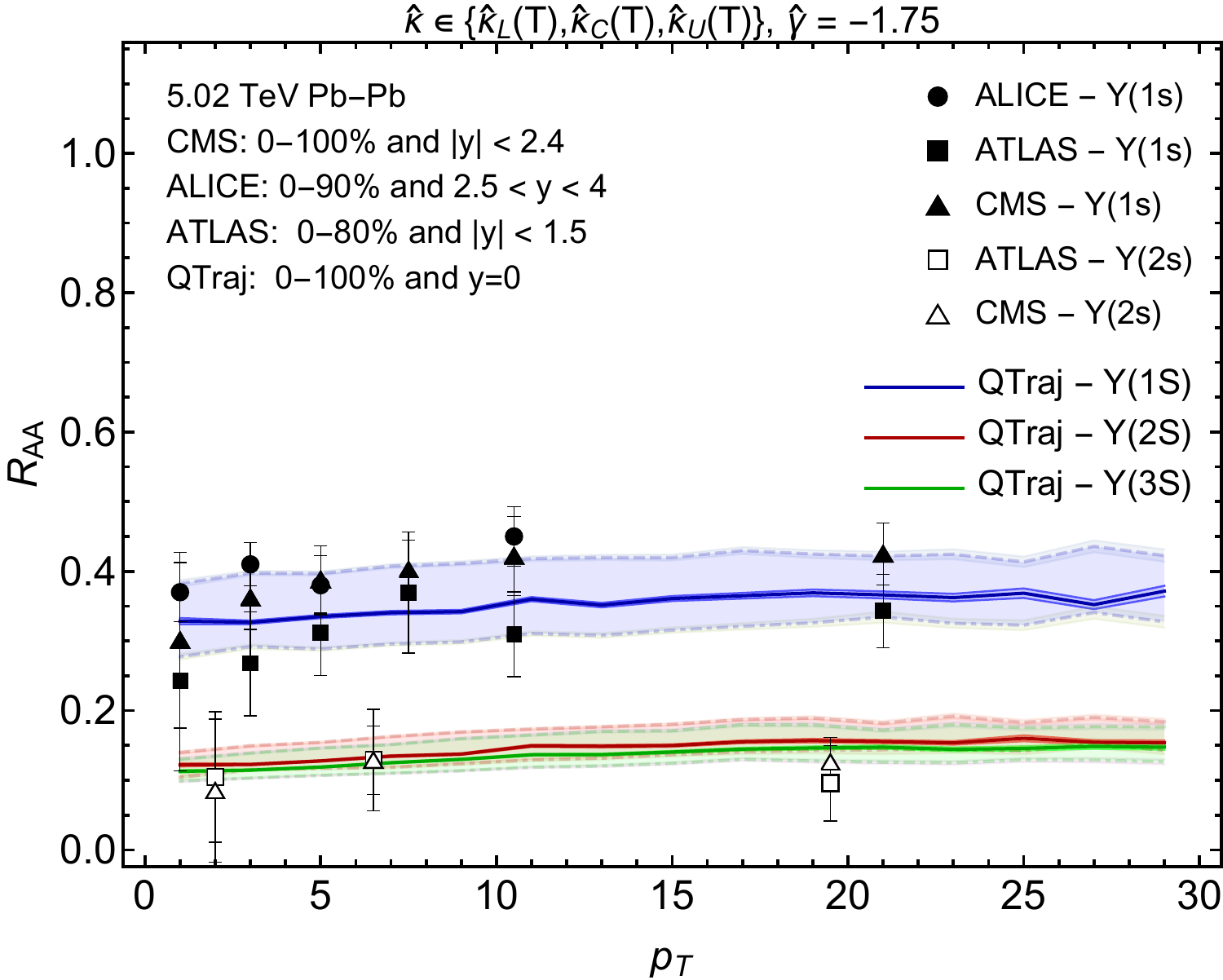}
}
\vspace{1mm}
\caption{
The nuclear modification factor $R_{AA}$ of the $\Upsilon(1S)$, $\Upsilon(2S)$, and $\Upsilon(3S)$ as a function of $N_{\text{part}}$ compared to experimental measurements from the ALICE~\cite{Acharya:2020kls}, ATLAS~\cite{ATLAS5TeV}, and CMS~\cite{Sirunyan:2018nsz} collaborations.
The bands indicate variation with respect to $\hat{\kappa}(T)$ (left) and $\hat{\gamma}$ (right).
The central curves represent the central values of $\hat{\kappa}(T)$ and $\hat{\gamma}$, and the dashed and dot-dashed lines represent the lower and upper values, respectively, of $\hat{\kappa}(T)$ and $\hat{\gamma}$. 		
}
\label{fig:raa}
\end{figure}
%%%%%%%%%%%%%%%%%%%%%%%%%%%%%%%%%%%%%%%%%%%%%%%%%%%%%%%%%%%

To compute the nuclear suppression in $AA$ collisions we compute the survival probability for a large number of physical trajectories with Monte-Carlo-sampled initial production points and transverse momenta.  Along each physical trajectory we additionally average over a set of quantum trajectories in which different quantum evolutions are sampled.  Once the survival probability for each state under consideration is computed, we then perform late-time excited state feed down using a feed down matrix $F$ constructed from the measured branching ratios and cross sections for bottomonium states~\cite{Brambilla:2020qwo}.  

With this one can compute the nuclear suppression of state $i$ from
\be
R^{i}_{AA}(c,p_T,\phi) = \frac{\left(F \cdot S(c,p_T,\phi) \cdot \vec{\sigma}_{\text{direct}}\right)^{i}}{\vec{\sigma}_{\text{exp}}^{i}} \, ,
\label{eq:feeddown}
\ee
where $ \vec{\sigma}_{\text{direct}}$ and $\vec{\sigma}_{\text{exp}}$ are the pre-feeddown and experimentally observed $pp$ cross sections for bottomonium production, respectively.  In this expression $c$, $p_T$, and $\phi$ correspond to the centrality, transverse momentum, and azimuthal angle bins being considered.  The survival probability $S(c,p_T,\phi)$ is a diagonal matrix containing each state's survival probablity along the diagonal.  For the background evolution, we use a three-dimensional anisotropic hydrodynamics code (aHydro3p1) that has been tuned to soft-observables such as the pion, kaon, and proton spectra and elliptic flow \cite{Alqahtani:2020paa}.  The aHydro3p1 code implements quasiparticle anisotropic hydrodynamics, including a lattice-based equation of state \cite{Martinez:2010sc,Florkowski:2010cf,Alqahtani:2015qja,Alqahtani:2017mhy}.

In practice we averaged over millions of sampled physical trajectories using a large one-dimensional lattice to solve for the stochastic evolution.  For the precise lattice spacing, time steps, initial conditions used, etc., we refer the reader to Ref.~\cite{Brambilla:2021wkt}.  In the left panel of Fig.~\ref{fig:raa} we plot the nuclear suppression factor as a function of the number or participants.  The band in the figure indicates our theoretical uncertainty in $R_{AA}$ stemming from the uncertainty in the transport coefficient $\kappa$.  Similar results are obtained when varying $\gamma$, however, the band is somewhat larger (see Ref.~\cite{Brambilla:2021wkt}).  As can be seen from  Fig.~\ref{fig:raa}, our predictions for the dependence of $R_{AA}$ on $N_\text{part}$ agree quite well with the reported experimental data.  In the right panel of Fig.~\ref{fig:raa} we present our predictions for the $p_T$-dependence of $R_{AA}$ and compare to experimental data.  Once again we see good agreement given current theoretical and experimental uncertainties.

%%%%%%%%%%%%%%%%%%%%%%%%%%%%%%%%%%%%%%%%%%%%%%%%%%%%
\begin{figure*}[t]
	\begin{center}
		\includegraphics[width=0.46\linewidth]{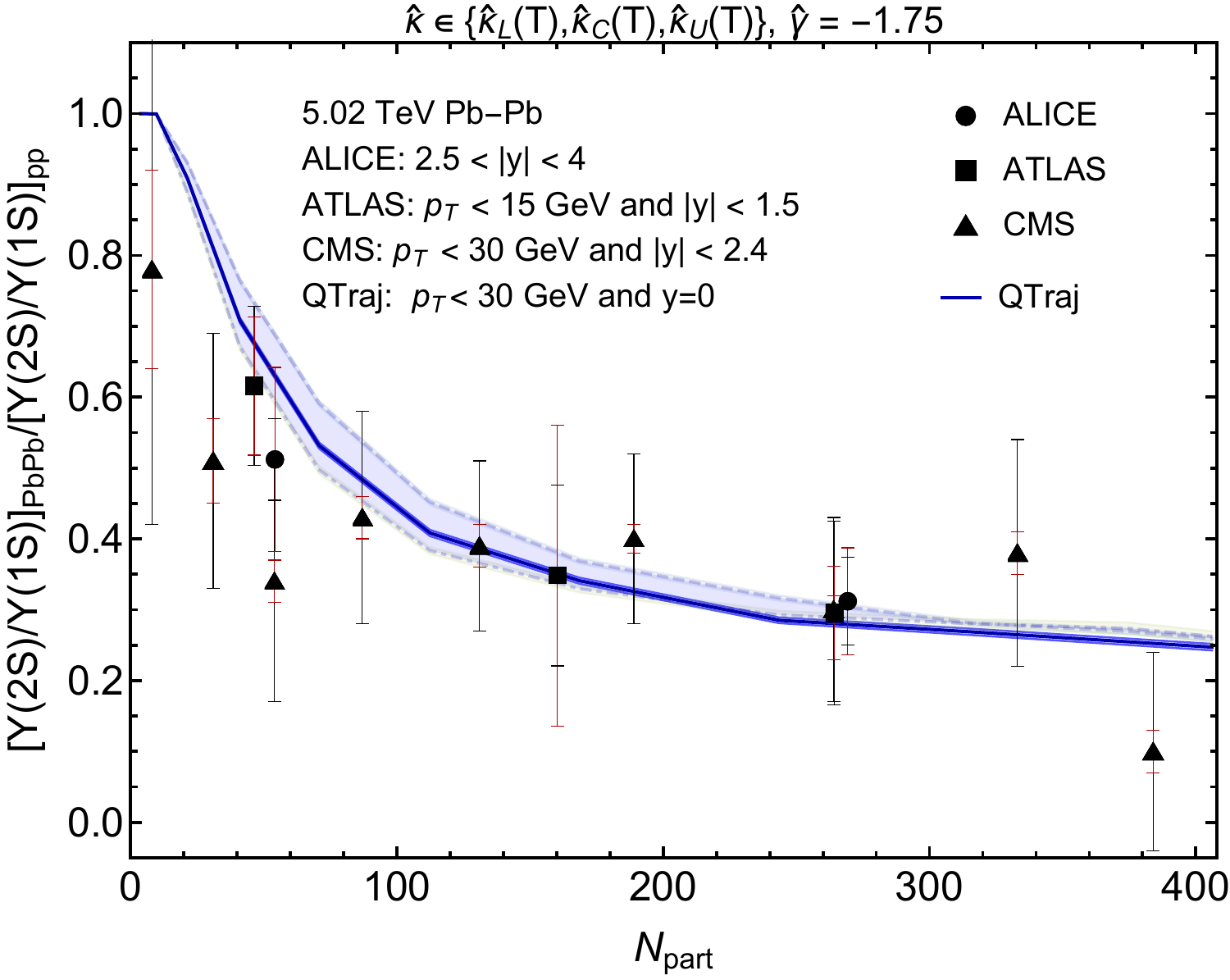} \hspace{4mm}
		\includegraphics[width=0.46\linewidth]{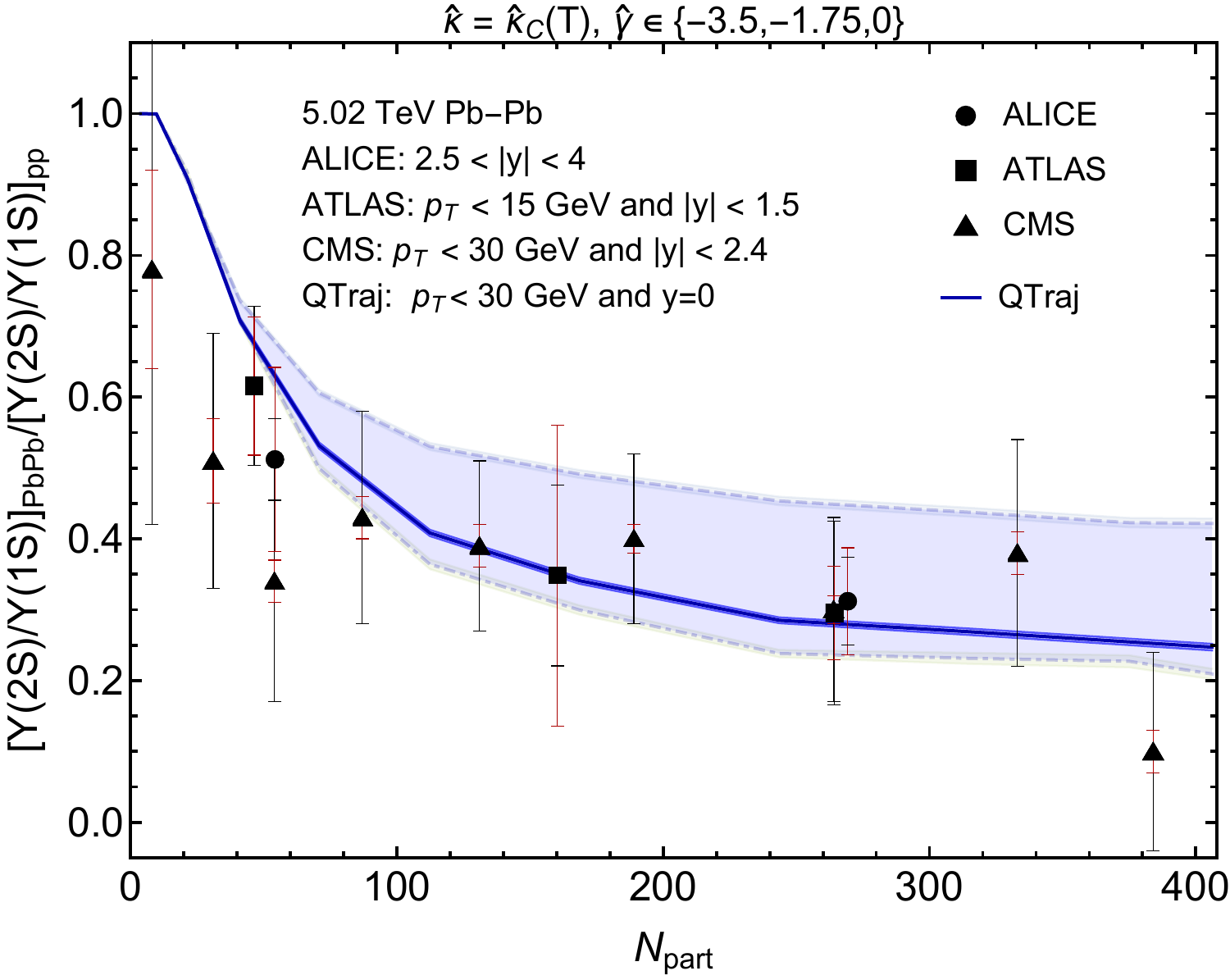}
	\end{center}
	\vspace{-5mm}
	\caption{(Color online)
		The double ratio of the nuclear modification factor $R_{AA}[\Upsilon(2S)]$ to $R_{AA}[\Upsilon(1S)]$ as a function of $N_{\text{part}}$ compared to experimental measurements of the ALICE~\cite{Acharya:2020kls}, ATLAS~\cite{ATLAS5TeV}, and CMS~\cite{CMS:2017ycw} collaborations.
		The bands indicate variation of $\hat{\kappa}(T)$ and $\hat{\gamma}$ as in Fig.~\ref{fig:raa}.
		The black and red experimental error bars represent statistical and systematic uncertainties, respectively.
	}
	\label{fig:2s_double_ratio_vs_npart}
\end{figure*}
%%%%%%%%%%%%%%%%%%%%%%%%%%%%%%%%%%%%%%%%%%%%%%%%%%%%

In Fig.~\ref{fig:2s_double_ratio_vs_npart} we present our predictions for the double ratio $[\Upsilon(2S)/\Upsilon(1S)]_\text{PbPb}/[\Upsilon(2S)/\Upsilon(1S)]_\text{pp}$.  In the left and right panels, the bands result from variation of $\kappa$ and $\gamma$, respectively.  As the left panel demonstrates, the dependence on $\kappa$ largely cancels when considering the double ratio; however, in the right panel we see that this ratio has a stronger dependence on $\gamma$.  Similar behavior was seen in the 3S to 1S double ratio \cite{Brambilla:2021wkt}.  This offers some hope that, with increased experimental statistics, one can use 2S/1S and 3S/1S double ratio data to constrain the transport coefficient $\gamma$.

%%%%%%%%%%%%%%%%%%%%%%%%%%%%%%%%%%%%%%%%%%%%%%%%%%%%
\begin{figure*}[t]
	\begin{center}
		\includegraphics[width=0.465\linewidth]{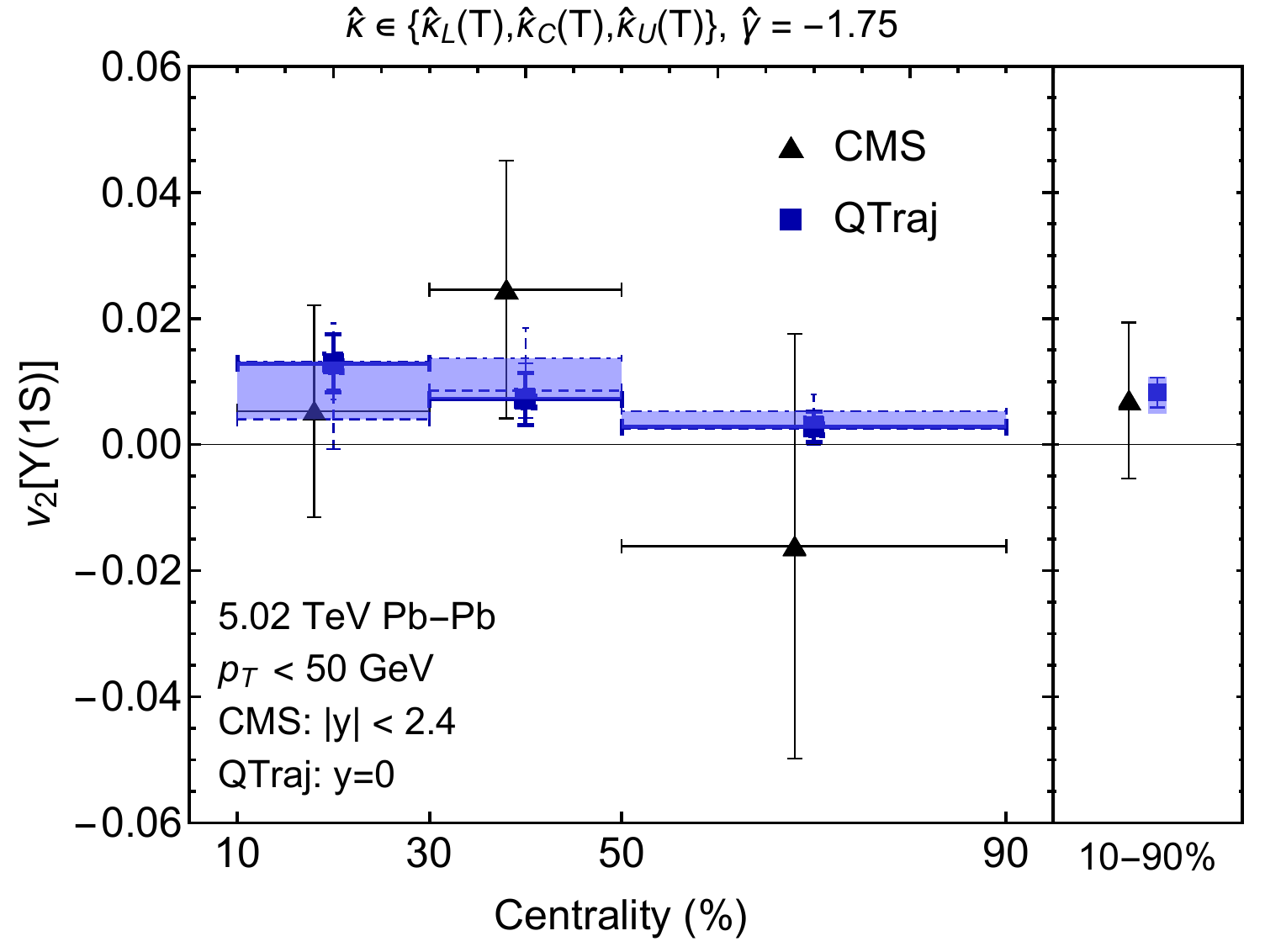} \hspace{2mm}
		\includegraphics[width=0.465\linewidth]{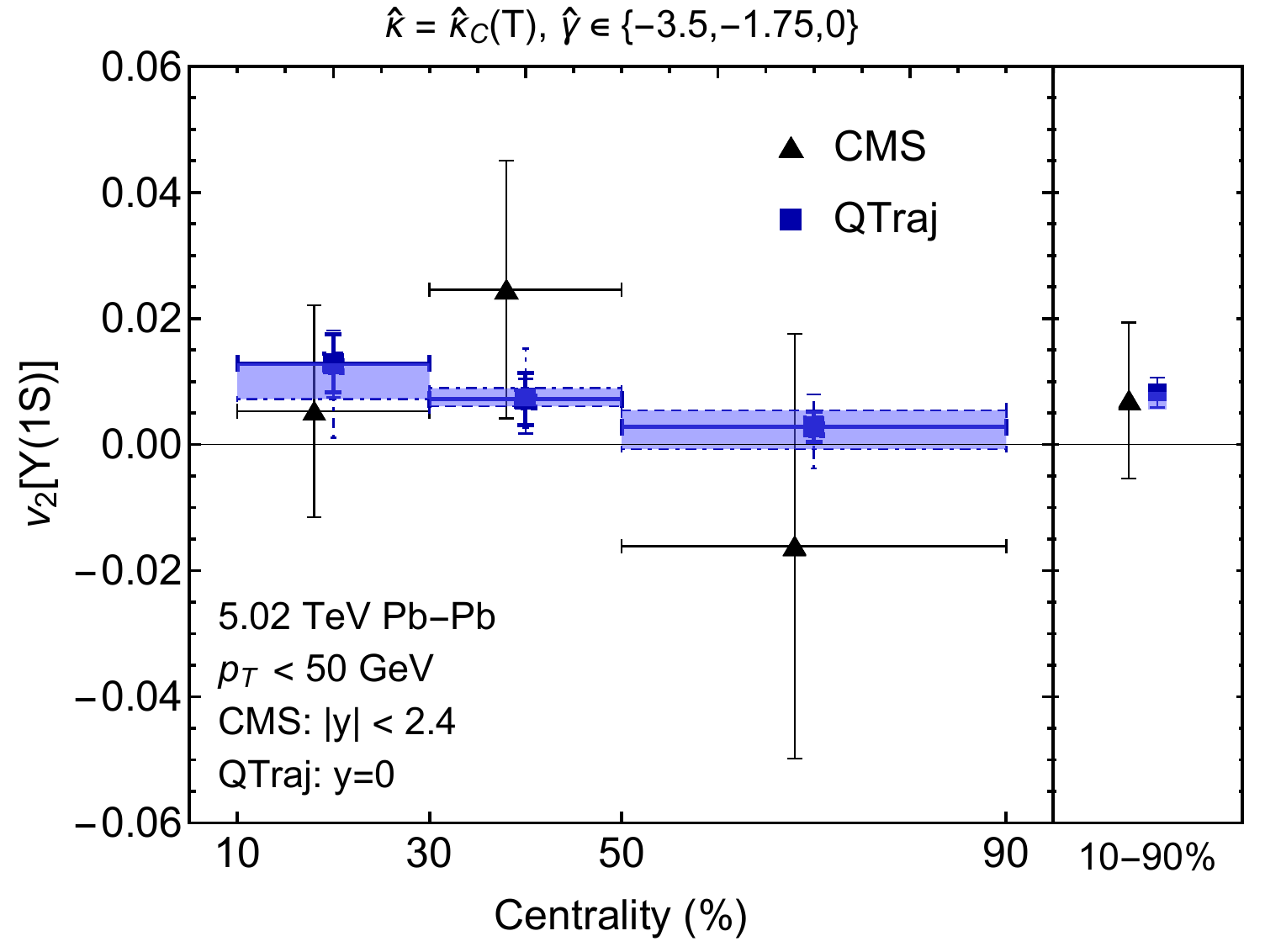}\;\;\;\;
	\end{center}
	\vspace{-5mm}
	\caption{(Color online)
		The elliptic flow $v_{2}$ of the $\Upsilon(1S)$ as a function of centrality compared to experimental measurements of the CMS~\cite{CMS:2020efs} collaboration.
		The QTraj bands represent uncertainties as in Fig.~\ref{fig:raa}.  QTraj error bars indicate the statistical uncertainty of our extraction.
	}
	\label{fig:v2_1S_vs_centrality}
\end{figure*}
%%%%%%%%%%%%%%%%%%%%%%%%%%%%%%%%%%%%%%%%%%%%%%%%%%%%

Finally, in Fig.~\ref{fig:v2_1S_vs_centrality} we present our predictions for the elliptic flow of the $\Upsilon(1S)$ in three different centrality classes along with a centrality-integrated measurement in the right subpanel of both figures. The left and right figures in Fig.~\ref{fig:v2_1S_vs_centrality} correspond to varying $\kappa$ and $\gamma$, respectively.  In the case of $v_2[\Upsilon(1S)]$, the variation with $\kappa$ and $\gamma$ are similar and our predictions are, again, in reasonable agreement with available data.  We note that if one focuses on the integrated results shown in the right sub-panel of each figure, our predictions have small systematic and statistical uncertainties and are consistent with CMS measurements within uncertainties.

\section{Conclusions and outlook}
\label{sec:conclusions}

In this proceedings contribution, I highlighted recent work that uses the framework of OQS together with pNRQCD to make phenomenological predictions for bottomonium suppression in the quark-gluon plasma.  The code used to generate the results presented herein has been released as an open-source package along with detailed documentation in Ref.~\cite{Omar:2021kra}.  Additional theory/data comparisons and discussions can be found in Ref.~\cite{Brambilla:2021wkt}.  Our results demonstrate that it is possible to understand experimental measurements of bottomonium suppression based on a first-principles approach that is fully quantum and incorporates features that are unique to non-abelian QCD, such as singlet-octet transitions.

Using pNRQCD and OQS methods, at leading-order one must solve an evolution equation of Lindblad form for the reduced density matrix.  Due to the difficulty of solving such a large matrix evolution equation, we made use of an algorithm originally developed for quantum optics applications, called the quantum trajectories algorithm.  With this method, solution of the 3D non-abelian Lindblad equation could be reduced to the solution of a 1D Schr\"odinger equation with a non-Hermitian Hamiltonian subject to stochastic internal transitions (angular momentum and color) dubbed quantum jumps.  This algorithm allows one to solve the 3D non-abelian Lindblad equation in a massively parallel manner due to the fact that each quantum trajectory is independent.

Looking to the future, it is necessary to include next-to-leading order corrections in the binding energy over the temperature, $E/T$, in order to assess the convergence of the high-temperature expansion used to obtain the leading-order Lindblad equation.  This will help to reduce the theoretical uncertainty in the low-temperature limit associated with the leading-order truncation and will be allow us to include drag/recoil effects.

\section*{Acknowledgments}
M.S. has been supported by the U.S. Department of Energy, Office of Science, Office of Nuclear Physics Award No.~DE-SC0013470.
M.S. also thanks the Ohio Supercomputer Center for support under the auspices of Project No.~PGS0253.  

%%%%%%%%%%%%%%%%%%%%%%%%%%%%%%%%%%%%%%%%%%%%%%%%%%%%%%%%%%%
\bibliography{strickland} 

\begin{thebibliography}{32}

\bibitem{Brambilla:2016wgg}
N.~Brambilla, M.A. Escobedo, J.~Soto, A.~Vairo, Phys. Rev. \textbf{D96}, 034021
  (2017), \texttt{1612.07248}

\bibitem{Brambilla:2017zei}
N.~Brambilla, M.A. Escobedo, J.~Soto, A.~Vairo, Phys. Rev. \textbf{D97}, 074009
  (2018), \texttt{1711.04515}

\bibitem{Brambilla:2020qwo}
N.~Brambilla, M.A. Escobedo, M.~Strickland, A.~Vairo, P.~Vander~Griend, J.H.
  Weber, JHEP \textbf{05}, 136 (2021), \texttt{2012.01240}

\bibitem{Brambilla:2021wkt}
N.~Brambilla, M.A. Escobedo, M.~Strickland, A.~Vairo, P.V. Griend, J.H. Weber
  (2021), \texttt{2107.06222}

\bibitem{Akamatsu:2020ypb}
Y.~Akamatsu (2020), \texttt{2009.10559}

\bibitem{Rothkopf:2020vfz}
A.~Rothkopf, Nucl. Phys. A \textbf{1005}, 121922 (2021), \texttt{2002.04938}

\bibitem{Yao:2021lus}
X.~Yao (2021), \texttt{2102.01736}

\bibitem{Omar:2021kra}
H.B. Omar, M.A. Escobedo, A.~Islam, M.~Strickland, S.~Thapa, P.V. Griend, J.H.
  Weber (2021), \texttt{2107.06147}

\bibitem{Dalibard:1992zz}
J.~Dalibard, Y.~Castin, K.~Molmer, Phys. Rev. Lett. \textbf{68}, 580 (1992)

\bibitem{Molmer:93}
K.~M{\o}lmer, Y.~Castin, J.~Dalibard, J. Opt. Soc. Am. B \textbf{10}, 524
  (1993)

\bibitem{Plenio:1997ep}
M.B. Plenio, P.L. Knight, Rev. Mod. Phys. \textbf{70}, 101 (1998),
  \texttt{quant-ph/9702007}

\bibitem{carmichael1999statistical}
H.~Carmichael, \emph{Statistical methods in quantum optics : master equations
  and fokker-planck equations} (Springer, New York, 1999), ISBN
  978-3-540-54882-9

\bibitem{weissbook}
U.~Weiss, \emph{Quantum Dissipative Systems} (WORLD SCIENTIFIC, 1993)

\bibitem{Daley:2014fha}
A.J. Daley, Adv. Phys. \textbf{63}, 77 (2014), \texttt{1405.6694}

\bibitem{Lindblad:1975ef}
G.~Lindblad, Commun. Math. Phys. \textbf{48}, 119 (1976)

\bibitem{Gorini:1975nb}
V.~Gorini, A.~Kossakowski, E.C.G. Sudarshan, J. Math. Phys. \textbf{17}, 821
  (1976)

\bibitem{Brambilla:2020siz}
N.~Brambilla, V.~Leino, P.~Petreczky, A.~Vairo, Phys. Rev. D \textbf{102},
  074503 (2020), \texttt{2007.10078}

\bibitem{Brambilla:2019tpt}
N.~Brambilla, M.A. Escobedo, A.~Vairo, P.~Vander~Griend, Phys. Rev. D
  \textbf{100}, 054025 (2019), \texttt{1903.08063}

\bibitem{Kim:2018yhk}
S.~Kim, P.~Petreczky, A.~Rothkopf, JHEP \textbf{11}, 088 (2018),
  \texttt{1808.08781}

\bibitem{Aarts:2011sm}
G.~Aarts, C.~Allton, S.~Kim, M.P. Lombardo, M.B. Oktay, S.M. Ryan, D.K.
  Sinclair, J.I. Skullerud, JHEP \textbf{11}, 103 (2011), \texttt{1109.4496}

\bibitem{Larsen:2019bwy}
R.~Larsen, S.~Meinel, S.~Mukherjee, P.~Petreczky, Phys. Rev. D \textbf{100},
  074506 (2019), \texttt{1908.08437}

\bibitem{Shi:2021qri}
S.~Shi, K.~Zhou, J.~Zhao, S.~Mukherjee, P.~Zhuang (2021), \texttt{2105.07862}

\bibitem{Acharya:2020kls}
S.~Acharya et~al. (ALICE) (2020), \texttt{2011.05758}

\bibitem{ATLAS5TeV}
{Songkyo Lee (ATLAS Collaboration)}, \emph{{Quarkonium production in Pb+Pb
  collisions with ATLAS}}, Quark Matter 2020
  \url{https://indico.cern.ch/event/792436/contributions/3535775/} (2017)

\bibitem{Sirunyan:2018nsz}
A.M. Sirunyan et~al. (CMS), Phys. Lett. B \textbf{790}, 270 (2019),
  \texttt{1805.09215}

\bibitem{Alqahtani:2020paa}
M.~Alqahtani, M.~Strickland (2020), \texttt{2008.07657}

\bibitem{Martinez:2010sc}
M.~Martinez, M.~Strickland, Nucl. Phys. \textbf{A848}, 183 (2010),
  \texttt{1007.0889}

\bibitem{Florkowski:2010cf}
W.~Florkowski, R.~Ryblewski, Phys.Rev. \textbf{C83}, 034907 (2011),
  \texttt{1007.0130}

\bibitem{Alqahtani:2015qja}
M.~Alqahtani, M.~Nopoush, M.~Strickland, Phys. Rev. \textbf{C92}, 054910
  (2015), \texttt{1509.02913}

\bibitem{Alqahtani:2017mhy}
M.~Alqahtani, M.~Nopoush, M.~Strickland, Prog. Part. Nucl. Phys. \textbf{101},
  204 (2018), \texttt{1712.03282}

\bibitem{CMS:2017ycw}
A.M. Sirunyan et~al. (CMS), Phys. Rev. Lett. \textbf{120}, 142301 (2018),
  \texttt{1706.05984}

\bibitem{CMS:2020efs}
A.M. Sirunyan et~al. (CMS), Phys. Lett. B \textbf{819}, 136385 (2021),
  \texttt{2006.07707}

\end{thebibliography}
%%%%%%%%%%%%%%%%%%%%%%%%%%%%%%%%%%%%%%%%%%%%%%%%%%%%%%%%%%%

\end{document}